# Policy choices and outcomes for offshore wind auctions globally

Malte Jansen[1], Philipp Beiter[2], Iegor Riepin[3], Felix Müsgens[3], Victor Juarez Guajardo-Fajardo[4], Iain Staffell[1], Bernard Bulder[5], Lena Kitzing[4]

[1] Centre for Environmental Policy, Imperial College London, United Kingdom

[2] National Renewable Energy Laboratory (NREL), United States

[3] Chair of Energy Economics, Brandenburg University of Technology, Germany

[4] Technical University of Denmark, Denmark

[5] TNO, Netherlands

## Abstract

Offshore wind energy is rapidly expanding, facilitated largely through auctions run by governments. We provide a detailed quantified overview of applied auction schemes, including geographical spread, volumes, results, and design specifications. Our comprehensive global dataset reveals heterogeneous designs. Although most remuneration designs provide some form of revenue stabilisation, their specific instrument choices vary and include feed-in tariffs, one-sided and two-sided contracts for difference, mandated power purchase agreements, and mandated renewable energy certificates.

We review the schemes used in all eight major offshore wind jurisdictions across Europe, Asia, and North America and evaluate bids in their jurisdictional context. We analyse cost competitiveness, likelihood of timely construction, occurrence of strategic bidding, and identify jurisdictional aspects that might have influenced auction results. We find that auctions are embedded within their respective regulatory and market design context, and are remarkably diverse, though with regional similarities. Auctions in each jurisdiction have evolved and tend to become more exposed to market price risks over time. Less mature markets are more prone to make use of lower-risk designs. Still, some form of revenue stabilisation is employed for all auctioned offshore wind energy farms analysed here, regardless of the specific policy choices. Our data confirm a coincidence of declining costs and growing diffusion of auction regimes.

# Highlights

- Global stock-taking of offshore wind auctions through statistical analysis.
- Auctions expected to dominate procurement in future, with estimated share of 97%
- Describing auction designs and outcomes for most of the 53.4 GW auctioned to date.
- Auction designs show a wide range of policy choices, embedded in regional contexts.
- Revenue stabilisation is a key procurement feature, especially in nascent markets.

# 1  Introduction

Offshore wind energy plays a critical role in meeting carbon-reduction goals and delivering an economic solution to power markets globally (International Renewable Energy Agency [IRENA], 2019a). The rollout of offshore wind energy is increasingly facilitated by competitive allocation mechanisms, commonly referred to as auctions. Auctions describe an economic mechanism to allocate goods or services with price formation governed through a bidding procedure. Here, we use the term broadly to encompass (seabed and remuneration) auctions as well as award mechanisms that – in a strict sense – are competitive procurement award mechanisms. Government entities conduct auctions to award the right to build and operate an offshore wind farm and to determine the support to be provided, replacing administrative allocations of permits and support used previously. This work is motivated by the fact that auctions have gained popularity with governments in Europe. In the United States and China, offshore wind energy is mainly realised through competitive solicitation for long-term contracts with public utilities, which we here also classify under the auction umbrella (IRENA, 2019b).

This trend towards auctions, and away from administratively determined price-controlled schemes (e.g., feed-in tariffs (FITs) or renewable obligations certificates [ROs]), represents a policy shift. It changes the risk exposure between asset owners and the government. In the offshore wind energy sector, auctions are emerging as the most prevalent form to realise new installations:  As of 2021, only 24% of all installed capacity was auctioned, but this share is expected to rise to 97% by 2030. Our comprehensive data analysis reveals that until 2017, 26 auctions were held for 41 wind farms with 16.8 GW of capacity, mostly in Europe. Between 2018 and 2021 alone, 32 auctions for 76 wind farms and 36.6 GW of capacity took place. At least 52 auctions are announced for the post-2021 period.

The increasingly dominant use of auctions merits a renewed and comprehensive inventory taking of market volume, implemented mechanisms, and assessment of policy design differences, all of which we present in this article. Previous studies on the topic have focused mainly on providing methodologies and tools for enhancing global comparability of mechanisms (though without analysing or assessing the different policy designs), or have focussed on few empirical country cases. Jansen et al. (2020) make offshore wind auction price outcomes in Europe comparable by introducing a methodology for harmonised expected revenues and differentiate in their analysis specific policy designs for Denmark, the United Kingdom (UK), the Netherlands, and Germany. Beiter et al. (2021) have produced calculation methodologies for different remuneration types for offshore wind energy utilised globally, to make value and revenue streams as a whole comparable. An early empirical study from Mast et al. (2007) compared experiences in Denmark and the UK, and describe offshore wind auction prospects for the Netherlands. DeCastro et al. (2019) describe (amongst others) the economic incentivisation of offshore wind energy in Europe, China, and the United States, and point to striking differences in policy approaches. Roberts (2020) sets the Chinese auctions into an international perspective. As methodology on assessing policy choices for auctions, we particularly draw on the work by Hochberg and Poudineh (2017), as well as Maurer and Barroso (2011), and empirically on work for Germany (Sach et al., 2019), the UK (Fitch-Roy and Woodman, 2016), Denmark (Garzón González and Kitzing, 2019), the United States (Beiter et al., 2020), Europe (Del Río et al., 2015), and globally (Ferroukhi et al., 2015; Mora et al., 2017).

The contribution of this paper is fourfold. First, we provide a comprehensive overview of all auctions held for offshore wind energy globally to date, alongside an open data base to facilitate data systematisation and transparency for a better understanding of auction mechanisms for offshore wind energy. Second, we systematically structure and describe key features of auction designs. Third, we contextualise results from auctions on a jurisdiction-specific level and analyse jurisdiction- and design-specific factors affecting the bids. Fourth, we make a cross-jurisdictional analysis on results and classify different auction designs, as well as commenting on policy implications on a global level, including trends towards subsidy-free offshore wind, the role of revenue stabilisation and interactions with seabed leases.

Our analysis shows that auctions are clearly the instrument of choice for deploying the vast majority of the future global offshore wind capacity of over 200 GW by 2030, regardless of each jurisdiction's individual policy setup. These auctions provide a varying degree of isolation from market price risks. The risk exposure varies based on the specific setup, rather than the type of instrument chosen, with a higher risk exposure chosen in more mature markets by policy makers. Our work has also laid bare the interactions between auctions and other policy processes on different levels of governance, e.g., seabed leasing, having implications on the bidding behaviour and auction outcomes. Together with a comprehensive data set, this paper provides the understanding on the nuances of auction designs for offshore wind energy for any interested party to analyse and compare across different continents and jurisdictions.

# 2 Approach

First, for the purpose of our comprehensive overview of the offshore wind energy market worldwide, we collect, compile and publish a publicly available dataset that includes all executed auctions for offshore wind energy to date. The content of the dataset was derived from Anatolitis and Roth (2020), Jansen et al. (2020), Zhao (2020), supplemented and updated with our own extensive research. We provide a dataset that is exceeding the current grey literature and proprietary databases significantly in breadth and depth, as visible in Supplementary Data 1. Our dataset contains all offshore wind farms from auctions through to the end of 2021, with geographic information (e.g., jurisdiction, province, city), developer name, execution status, tender capacity, and capacity of each wind energy project, winning bids, support scheme type, auction year and award date, final investment decision (FID) date, and construction and operation start—all indicated with their respective source. The dataset also captures planned auctions beyond 2021, with the caveat that these are likely to be incomplete. In Section 3.1, we show the entire dataset for all auctions worldwide, for all jurisdictions and all years, including announced auctions in the future. In Section 3.2, we focus on results from the most developed markets in eight jurisdictions, capturing 86.5% of all auctions globally to date by capacity. The full list is available in Supplementary Data 1. Our dataset covers 113 auctioned wind farms in 56 rounds, totalling 53.4 GW of awarded capacity, starting with the first occurrence in 2005 and covering the period until 2021. Additionally, we have identified 37 auctions planned for 2022 and beyond with 63.6 GW of capacity.

Second, we structure and describe key features of auction designs (Section 4) for the eight most developed markets. We present all jurisdictions that had auctioned a capacity of at least 1 GW installed at the end of 2021 and had at least two auctions —the minimum to establish a trend. Further, we only consider countries which had at least one of their projects mature to financial close (FID), indicating a jurisdiction's auction scheme ability to produce viable projects. We discuss the type of support mechanism, which can take the form of contract for differences (CfDs), FITs, mandated power purchase agreements (PPAs) for a predetermined capacity, or mandated offshore wind renewable energy certificates (ORECs). We further elaborate on the following design features: (1) support duration, (2) market reference prices, (3) inflation adjustment and indexation, (4) grid-connection cost allocation, (5) site development cost allocation, (6) penalty for noncompliance, and (7) frequency of auctions. We also comment on whether seabed leases are allocated separately. The categorisation builds and expands on earlier by Fitch-Roy (2015). Definitions of support mechanism and examples on items (1)–(7) are also provided in Supplementary Note 1.

Third, we analyse and contextualise the auctions results for offshore wind energy in each jurisdiction and discuss jurisdiction- and design-specific factors affecting the bids (Section 4). For this contextualisation exercise, we gather expertise from eight economists and offshore wind energy industry experts from all relevant jurisdictions. Our discussion includes the following aspects: (a) project realisations observed to date, (b) likelihood of "option bidding" with a risk of nondelivery of projects, (c) strategic bidding that might lead to a deviation from the (presumed) underlying costs of the bids, and (d) other political or economic factors relevant to specific jurisdictions.

Fourth, we make a cross-jurisdictional analysis and classification of auction mechanisms according to their risk/market relation, and compare some of the most striking designs and results, including multiple zero bids and lotteries, as well as interactions between seabed lease auctions and support auctions. Following this, we provide an overarching discussion that focuses on policy implications (Section 0). We shed light on the current transition from "procuring a costly service" to "awarding a valuable right" to build an offshore wind project. While this is a positive development overall, it poses a few new questions for the design of auctions.

# 3 Worldwide offshore wind energy auctions

We consider a large variety of different competitive allocation processes under the term "auction". Whilst this includes price-competitive bidding in some form in every jurisdiction investigated here, it stretches from the right to obtain support payments (e.g., CfDs or competitive allocation of PPAs) to the access to valuable wind resources (e.g., via seabed leases), using a multitude of different policy tools. In some cases, support auctions are separate from seabed lease activities (e.g. United Kingdom, United States), whereas in other cases they are not (Netherlands, Germany). Whilst auction prices determine the right to build (in most cases through the so-called strike price), they may not be the only criteria for a bidder being awarded, but an emphasis on delivery and the technical capabilities is possible. Following our global stock-take exercise for all markets with auctions (Section 3.1), we explore the most developed offshore wind markets (Section 3.2) and discuss their specific policy design and differences (Section 4).

## 3.1 Comprehensive worldwide overview

In the world, 53.4 GW of offshore wind energy have been awarded based on auctions at the time of writing, which is more than the 48.4 GW of installed capacity worldwide (World Forum Offshore Wind, 2021). The time lag between auction and commissioning date means that just 11.7 GW (22%) out of the overall auctioned capacity have been operational by 2021, roughly doubling yearly, from 5.3 GW commissioned in 2020 and 2.7 GW in 2019. We expect that 96% of all global offshore wind capacity by 2030 will have been auctioned, as most jurisdictions will use this policy approach exclusively.

Table 1 lists all global offshore wind energy markets with capacity installed or under construction as of 2021, expected capacity additions for 2022–2030, and the share of auctions for the installed capacity, broken down by jurisdictions and regions. Further, 68.2 GW of announced auctions from 2022 onwards are on record, which would more than double the current total global offshore wind capacity. This means that offshore wind energy is set to increase by 121.6 GW from auctions alone throughout the 2020s. The list of announced projects is likely incomplete and further additions and cancellations should be expected. The complete dataset is accessible in Supplementary Data 1. In addition, policy targets indicate that a total of 224 GW of offshore wind will be realised by 2030.

Table 1: Overview of jurisdictions with auctions for offshore wind energy analysed in this study. Unless indicated otherwise, data based on our own research are shown in Supplementary Data 1. Ordered by installed capacity end of 2021.

| Jurisdiction | Auctions | | Auctioned capacity procured until 2021 [in MW] | | | | | | Capacity from 2022 onward [in MW] | | | | | | Auction share | |
|---|---|---|---|---|---|---|---|---|---|---|---|---|---|---|---|---|
| | Number of auctions conducted (end 2021)[a] | Auctioned capacity (end of 2021) [in MW] | Total installed capacity end of 2021[b] | ...of which auctioned | Under construction (end 2021)[b] | ...of which auctioned | Future additions (projects announced + policy targets 2030/35)[c] | ...of which auctioned[d] | Under construction + announced future projects (developer + policy targets (2030/35) | ...of which auctioned | Expected additional auctions (2022–2030) | Total auctions (2022-2030) | Share in 2021 installed capacity | Share in future capacity |
| China | 7 | 5,850 | 19,747 | 3,000 | 7,992 | 2,850 | 58,391 | 4,650 | 66,383 | 7,500 | 53,741 | 59,591 | 15.2% | 92.3% |
| United Kingdom | 4 | 9,824[e] | 12,281 | 4,358 | 2,990 | 2,288 | 37,623 | 11,466 | 40,613 | 13,754 | 26,157 | 35,981 | 35.5% | 98.3% |
| Germany | 3 | 4,058 | 7,701 | 0 | 0 | 0 | 22,299 | 4,058 | 22,299 | 4,058 | 18,241 | 22,299 | 0.0% | 100% |
| Netherlands | 8 | 4,531 | 3,010 | 2,232 | 2,299 | 2,299 | 6,191 | 6,100 | 8,490 | 8,399 | 91 | 4,622 | 74.1% | 100% |
| Denmark | 8 | 3,156 | 2,343 | 1,806 | 0 | 0 | 7,350 | 1,350 | 7,350 | 1,350 | 6,000 | 9,156 | 77.1% | 100% |
| Belgium | 0 | 0 | 2,434 | 0 | 0 | 0 | 3,366 | 0 | 3,366 | 0 | 0 | - | 0.0% | 0% |
| Taiwan | 2[f] | 5,500 | 237 | 237 | 2,505 | 2,505 | 11,531 | 2,758 | 14,036 | 5,263 | 8,773 | 14,273 | 100.0% | 100% |
| South Korea | 0 | 0 | 104 | 0 | 0 | 0 | 8,200 | 0 | 8,200 | 0 | 8,200 | - | 0.0% | 100% |
| Japan | 3[g] | 1,532 | 85 | 0 | 140 | 0 | 9,915 | 1,532 | 10,055 | 1,532 | 8,383 | 9,915 | 0.0% | 98.6% |
| United States | 9[h] | 8,528 | 42 | 0 | 0 | 0 | 25,850 | 8,528 | 25,850 | 8,528 | 17,322 | 25,850 | 0.0% | 100% |
| Italy | 1 | 30 | 30 | 30 | 30 | 30 | 900 | 0 | 930 | 30 | 900 | - | 100.0% | 100% |
| Norway | 0 | 0 | 6 | 0 | 88 | 0 | 4,500 | 0 | 4,588 | 0 | 4,500 | - | 0.0% | 98.1% |
| France | 5 | 3,679 | 2 | 0 | 976 | 976 | 4,224 | 2,703 | 5,200 | 3,679 | 1,521 | 5,200 | 0.0% | 100% |
| India | 1[i] | 0 | 0 | 0 | 0 | 0 | 4,000 | 0 | 4,000 | 0 | 4,000 | - | - | 100% |
| Poland | 1 | 5,490 | 0 | 0 | 0 | 0 | 10,900 | 5,400 | 10,900 | 5,400 | 5,500 | 10,990 | - | 100% |
| Rest of world | 0 | 0 | 416 | 0 | 0 | 0 | n/a | 0 | n/a | 0 | 0 | - | 0.0% | n/a |
| **Europe** | 30 | 30,768 | 27,807 | 8,426 | 6,383 | 5,593 | 97,353 | 31,077 | 103,736 | 36,670 | 66,060 | 88,248 | 30.3% | 99.0% |
| **Asia** | 13 | 12,882 | 20,173 | 3,237 | 10,637 | 5,355 | 92,037 | 8,940 | 102,674 | 14,295 | 83,097 | 83,779 | 16.0% | 94.9% |
| **North America** | 9 | 8,528 | 42 | 0 | 0 | 0 | 25,850 | 8,528 | 25,850 | 8,528 | 17,322 | 25,850 | 0.0% | 100.0% |
| **Total** | 52 | 52,178 | 48,438 | 11,663 | 17,020 | 10,948 | 215,240 | 48,545 | 232,260 | 59,493 | 166,479 | 197,877 | **24.1%** | **97.3%** |

[a] *Indicates that tendering process has started (as opposed to finished with results).*

[b] *Source: Statistics from World Forum Offshore Wind (World Forum Offshore Wind, 2021).*

[c] *Sources: NREL 2020 Offshore Wind Technology Data Update (forthcoming) for United Kingdom, China, Taiwan, United States, Norway, and South Korea. Numbers for Denmark updated based on our own research. All other numbers are forthcoming capacity additions based on International Energy Agency Offshore Wind Outlook 2019 (International Energy Agency, 2019), with installed capacity and capacity under construction deducted.*

[d] *Indicates that capacity is already auctioned and to be constructed from 2022 onward.*

[e] *Seagreen wind farm is support through a CfD supported with capacity of 454 MW. The entire wind farm capacity is 1,075 MW, which means that a 621 MW share is operated on a merchant basis and thus not included here.*

[f] *One auction round on price, one auction round with administrative FIT.*

[g] *Includes Happo-Noshiro zone wind farm auction, which has started but not concluded by the time of writing. Not considered in the finally auctioned capacity yet.*

[h] *The auction 'Request for Proposals (RFP) for 2 GW offshore wind' in Massachusetts, USA, has produced wind farms submitting proposals at different times in the years 2018 and 2019 and are thus counted as a separate auction.*

[i] *1,000-MW tender process for Gujarat offshore wind farm started, but currently postponed (REVE, 2021).*

While most initial auctions occurred in Europe (31.8 GW), the United States (8.5 GW) and Asian jurisdictions (13.1 GW) have recently added large amounts of capacity. So far, 2019 holds the record with 14.0 GW of capacity in 11 auction rounds, with 2020 seeing only 3.7 GW in five rounds, with only one auction in the two most mature markets of China and the UK. The COVID-19 pandemic has postponed some activity in China (Power Engineering International, 2021), though in the UK, no auctions were planned for that year. For 2021, our dataset contains 7.9 GW in four auctions. Delays in auction execution and award means that auctions initially planned for 2021 have ticked over the next year. Now, 2022 shapes up to be setting records with 17 auctions for 28.1 GW of capacity.

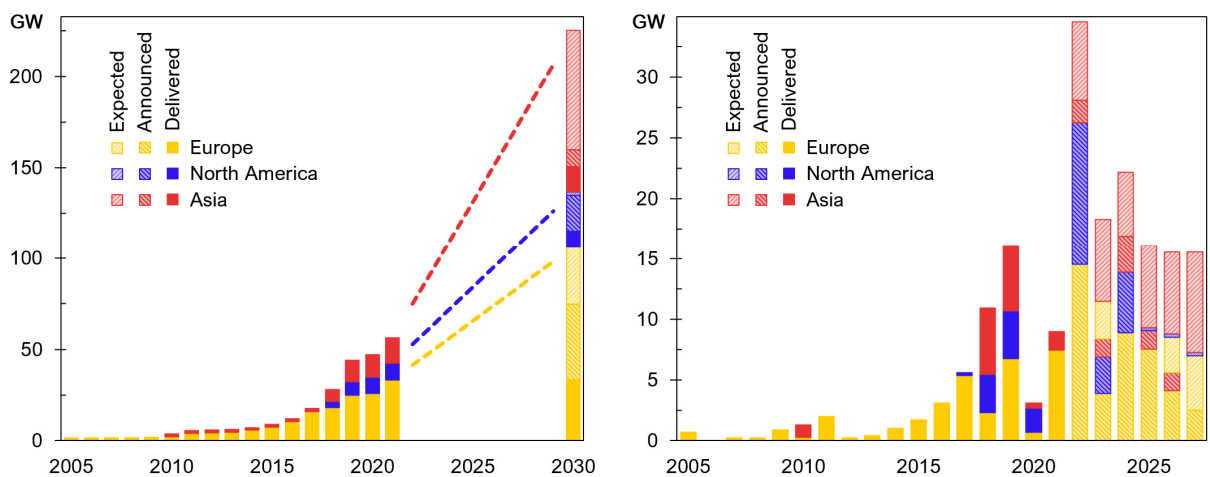

*Figure 1: Awarded capacity of offshore wind farms by region. Left panel shows the cumulative total, right panel shows the annual additions. Auctions from 2022 onward are either already announced but yet to be executed (darker hatched), or yet to be announced but are expected based on policy targets (lighter hatched). Expected capacity is equally distributed across the future years to give a sense of scale, not to indicate the actual timing of future auction rounds.*

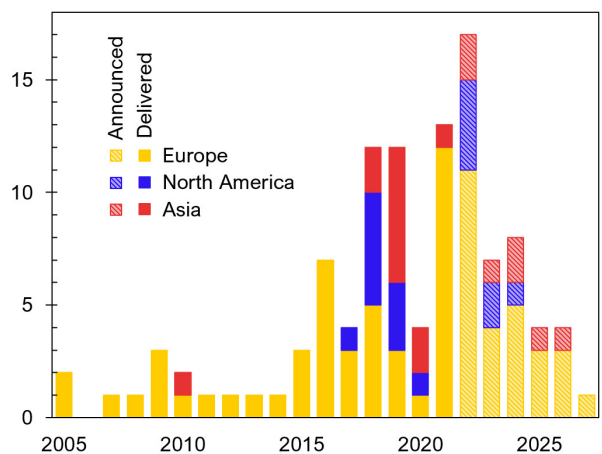

*Figure 2: Number of auctions held by region. Auctions from 2022 onward are announced, but yet to be executed (shown as hatched).*

## 3.2 Analysis of focus jurisdictions

In this study, we focus on jurisdictions (i.e., countries, regions, or federal states) with an auction mechanism implemented by the end of 2021, more than 1 GW already auctioned, at least two rounds and at least one project with financial close. These jurisdictions are China, the United Kingdom, Germany, the Netherlands, Denmark, Taiwan, the United States and France, in descending order of their installed capacity.

Auctions are still primarily used to allocate support through different remuneration scheme design. In our focus jurisdictions, five different remuneration schemes are used. Almost half of the capacity, 26.5 GW (51.7%) is supported through CfDs, with 17.9 GW using two-sided (34.9%) and 8.6 GW a one-sided CfD (16.8%). A further 9.6 GW were given competitively allocated FITs (18.8%), and 4.8 GW were auctioned but received an administratively determined FIT. In total, 5.3 GW (10.3%) are based on certificates (ORECs), and 5.0 GW on PPAs (9.7%). Earlier auctions were mainly based on competitively allocated feed-in tariffs in China and CfDs in Europe. The diversity of support schemes is increasing, as more jurisdictions are supporting offshore wind energy. We discuss the implications of the different funding schemes after the jurisdiction chapters in Section 0. Despite record-breaking wind farms sizes now routinely exceeding 1 GW, the average size of auctioned wind farms has remained relatively stable around 500 MW over the last 5 years.

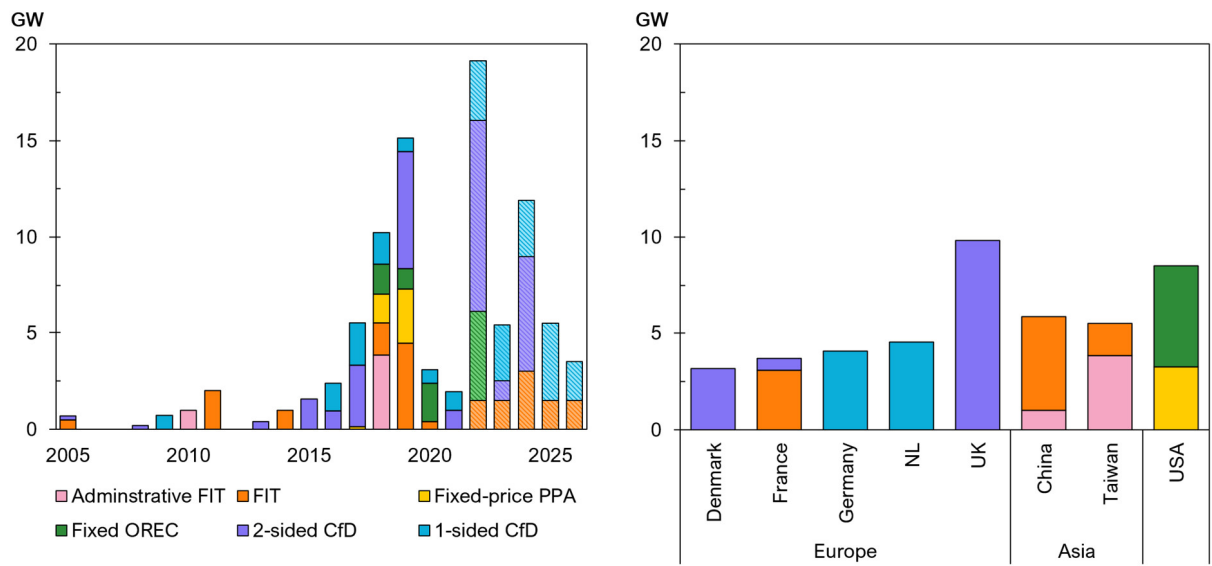

*Figure 3: Capacity of offshore wind from auctions. Capacity is broken down by year and mechanism on the left panel. The hatched areas on the right panel indicated announced auctions that are yet to be executed. The right panel shows the breakdown by jurisdiction and mechanism, showing only capacity that is already delivered.*

Figure 3 shows early wind farms predominantly received fixed-price support, such as FiTs and green certificates, which were superseded by one-sided and two-sided CfDs mainly in Europe. New entrants in the United States and Asia are relying mainly on fixed-price support, such as FITs or PPAs. As auctioned capacity is increasing, the awarded raw bid prices are falling. Notably, several wind farms represented here will be subsidy-free while formally attributed to a funding scheme. This is particularly relevant for Germany, the Netherlands, Denmark and, in essence, the UK (Jansen et al., 2020). Figure 4 compares the prices between different schemes. Based on the raw bid prices, we can observe that prices are falling, apart from administrative FITs. This can be explained by their use in Asia, which is a less mature market.

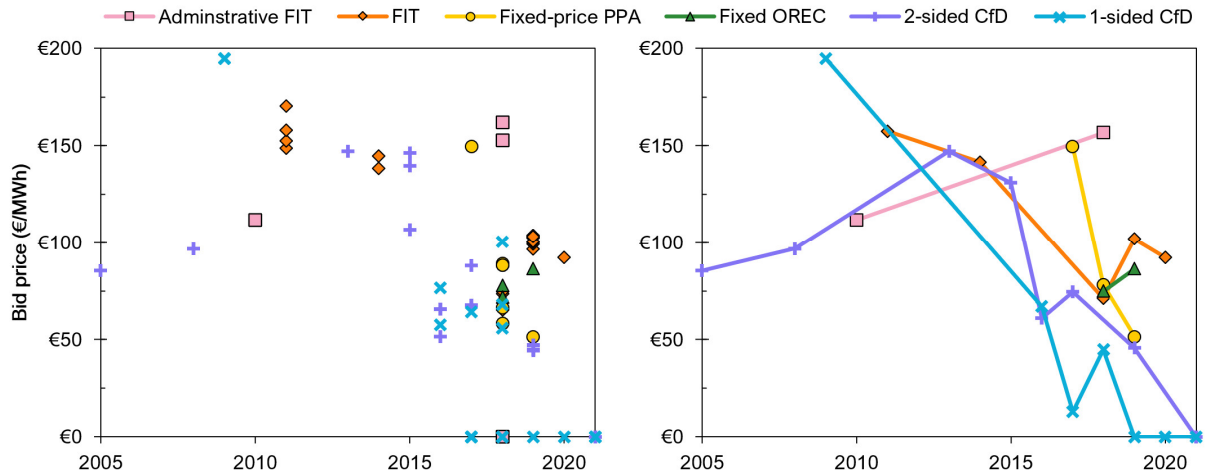

*Figure 4: Awarded prices for offshore wind farms in the eight jurisdictions we cover, categorised by the auction mechanism. Individual raw bids are shown in the left panel, and the average bids per mechanism are shown in the right panel. The awarded prices broken down by regions are in Supplementary Figure 1. Note that the bid levels cannot be compared directly, unless harmonised (e.g., see (Beiter et al., 2021; Jansen et al., 2020)), as they do not correspond to the full revenues of the wind farms over their lifetime.*

Table 2 summarises our detailed findings on different auction schemes. From this we can establish that the UK, China, Netherlands and Denmark are the clear leaders in realised offshore wind capacity from auctions, with very different funding approaches. China has the largest capacity from auctions under construction, followed closely by the UK, with its tight delivery deadlines and 99.9% of capacity with FID, and the highest capacity from auctions currently under construction. The share of FID taken is also high in the Netherlands and Denmark. It is about half of this in China, Taiwan, and France. All other wind farms were still awaiting FID at the time of writing.

*Table 2: Overview of auction outcomes for focus jurisdictions in this study. Raw data available in Supplementary Data 1. Contextualisation of results with analysis for each jurisdiction given in the subsequent Section 4. Numbers are valid for the end of 2021.*

| Jurisdiction | Number of auctions | Number of wind farms | ...of which had financial close (in % of capacity)[a] | Capacity-weighted average prices (min, max) in local currency /MWh at award time)[b] | Capacity-weighted average prices (min, max) in €$_{2020}$ /MWh | Years of auctions (post 2021 are announcements) | Grid connection responsibility | Funding mechanism (CfD, FIT, tax credit) | Indexation to inflation | Site development | Seabed lease auction | Penalties for non-compliance |
|---|---|---|---|---|---|---|---|---|---|---|---|---|
| China | 7 | 21 | 9[c] (36.8%) | ¥ 760.9 (684.0, 795.0) | 103.1 (92.3, 111.8) | 2010, 2019, 2020 | n/a | Administrative FIT/ competitive FIT | No | n/a | n/a | Permit loss w/o construction > 2 years |
| United Kingdom | 4 | 9 | 8 (99.9%) | £$_{2012}$ 56.8[d] (39.7, 119.9) | 66.5 (44.9, 146.4) | 2015, 2017, 2019, 2021, 2023, 2024 | Bidder | Two-sided CfD | Yes | Bidder | Yes, separate Crown Estate auction prior CfD | Non-delivery: Banned for 2 years |
| Germany | 3 | 13 | 9 (54.2%) | € 14.7 (0, 98.3) | 15.1 (0.0, 100.5) | 2017, 2018, 2021-2023 | Socialised | One-sided CfD | No | Society | No | Financial €0.1-0.2/MW |
| Netherlands | 8 | 8 | 7 (83.2%) | €25.4 (0, 76.6) | 26.8 (0.0, 76.6) | 2016-2022, 2026 | Socialised | One-sided CfD | No | Society | Yes, part of auction criteria | Non-d.: €10m Late:€3.5m/mo |
| Denmark | 8 | 8 | 5 (83.8%) | DKK 627.0 (0.01[e], 1051.0) | 89.4 (51.5, 147.3) | 2005, 2008, 2013, 2015, 2016, 2021, 2022, 2024 | Socialised (until '16)/Bidder (from '21) | Two-sided CfD | No | Bidder | No | Non-d./Late: ≤€m0.15/MW+ less supported production |
| Taiwan | 2[f] | 14 | 6 (38.2%) | NT$ 2,488.9 (2,224.5, 5,516.0) | 90.8 (65.3, 161.9) | 2018, 2021-2023 | Socialised | Administrative FIT/competitive FIT | No | Society | No | - |
| United States | 9 | 13 | 1 (9.4%) | US$ 84.6 (58.5, 163.0) | 75.5 (51.5, 149.8) | 2017-2021 | Bidder | Fixed OREC, fixed-priced PPA | Yes | Bidder | Yes, prior federal seabed auctions | - |
| France | 5 | 8 | 4 (42.0%) | €123.7 (44.0, 155.0) | 133.7 (44.3, 170.5) | 2005, 2011, 2014, 2019, 2021-2024 | Socialised[g] | FIT, two-sided CfD | Yes | Bidder | No | CfD shorted by the number of days delayed |

[a] *Up to date as of December 31st, 2021.*

[b] *Nominal values for renewable support schemes without indexation of payments to inflation. These prices typically indicate nominal strike prices of an auction in Europe and Asia, whereas in the United States they are typically the "levelized nominal price" of a PPA or OREC; payments in the UK, United States, and France are indexed to inflation, although only the UK has a common base year for inflation.*

[c] *Information on financial close is not available explicitly. We are assuming that all wind farms with a commissioning date up to 2021 have reached financial close.*

[d] *All bids are given in £$_{2012}$, as this is the basis for inflation adjustment.*

[e] *Thor two-sided CfD was awarded following a lottery, as several bidders had submitted the minimum bid prize. Overall payments by the developers is capped at DKK 2.8 bn., corresponding to MEUR 375 (2018 prices)*

[f] *The first round of 11 wind farms and 3.8 GW were auctions with an administrative FIT. Second round for three wind farms of 1.7 GW in total was auctioned mainly on price. Bid prices include the price from both auctions.*

[g] *Initial two tenders included grid connection. Grid connection is not carried out by the grid operator after renegotiation on bid price.*

# 4 Auction designs

Here, we introduce auction schemes for offshore wind energy in each jurisdiction. The jurisdiction section is divided into: (1) auction mechanisms, covering legislation, pricing rules, and design; and (2) analysis of the bids observed, the likelihood of strategic bidding behaviour, and any jurisdiction-specific issues. Jurisdictions are presented in order of installed capacity (end of 2021), and then by auctioned capacity. The raw bids and their evolution over time are shown below.

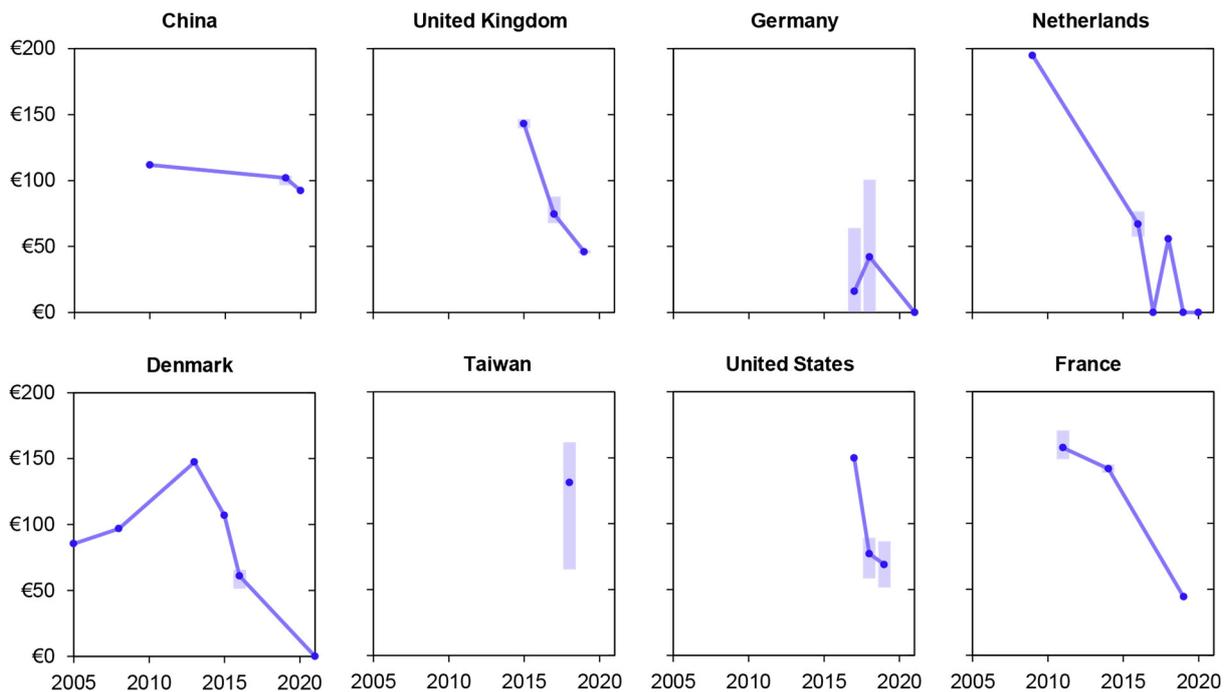

Figure 5: Raw bids from offshore wind auctions over time within each jurisdiction. Areas indicate the range of bids observed in one particular year, in which case the marker indicates the average price of bids in the year. Note that the bid in the Netherlands for 2018 is for a technology test site of 20 MW only.

## 4.1 China

### 4.1.1 Mechanisms

The 2005 Renewable Energy Law provides a legislative framework for developing renewable energy in China (International Energy Agency [IEA] and IRENA, 2017), stipulating a guaranteed purchase requirement with its price determined by the National Development and Reform

Commission (NDRC), but with details on auction design or FITs omitted. This has led to China rapidly growing to be the largest market globally (Lee and Zhao, 2020; World Forum Offshore Wind, 2021).

The first offshore wind concession tender in China dates back to 2010, organised by the National Energy Administration, for two nearshore projects (30 MW each) and two intertidal projects (20 MW each) (Chinese Wind Energy Association, 2011). The tender prioritises the bid closer to the average bid price rather than the lowest. The price component is accounting for 55% of the total score in the selection process. In June 2014, the NDRC decided to pause offshore wind energy auctions/tenders and introduced a FIT policy for offshore wind; CNY850/MWh (€108/MWh) for nearshore and CNY750/MWh (€95/MWh) for intertidal (Wei et al., 2021).

The National Energy Agency announced in May 2018 that all offshore wind energy projects approved from 2019 onward must compete in price-competitive auction regimes that are held by provincial governments, setting a price cap of CNY800/MWh and 750/MWh for 2019 and 2020, respectively (Wei et al., 2021). The winning bid price cannot exceed the stated FIT rate. Projects approved before the end of 2018 are still eligible for a FIT, if fully connected before the end of 2021. By the end of 2019, 20 offshore wind energy projects have gone through the auction process and only one wind farm was added in 2020 due to the coronavirus outbreak.

The first auction under the new regulation launched in June 2019 was for the Fengxian offshore wind project (200 MW) in Shanghai. The evaluation follows a scoring approach different from the earlier 2010 auction rounds, which includes scoring on enterprise capability (25%), progressiveness of technology (15%), plan of implementation (20%), and price (40%). The price criteria prioritises bids with the lowest rather than the average bid price, as was the case in previous auctions. Bids with the lowest price score highest (40 out of 100 points) and bids with higher prices score 0.2 to 0.3 points lower for every CNY10/MWh (€1.3/MWh) increment in price. The auction was won by Shanghai Electric Power and Shanghai Green Environmental Protection Energy jointly at CNY738.8/MWh (€93.42/MWh). Auctions in November/December 2019 include tenders for seven projects in the Zhejiang province (five in Wenzhou and two in Ningbo), four projects in Dalian, Liaoning province, and four projects in Yantai, Shandong province. The evaluation/selection criteria are similar to the Fengxian project in Shanghai, but weightings can vary in some cases as price scoring is a function of the difference between the bid and the FIT (e.g., the projects in Dalian). All the auctions have static sealed-bids and marginal pricing applied. Seabed leases are not allocated separately from the support auction. Project consent is revoked, if the project has not commenced offshore construction within two years after receiving approval (Hogan Lovells, 2020).

Offshore wind energy projects have guaranteed grid access in China. The two state grid companies, China State Grid and China Southern Power Grid, are responsible for the grid connection of offshore wind projects. However, it was claimed that the bottleneck on supply chain such as wind turbine blades, main bearings, and offshore cables, as well as turbine installation vessels, could limit the potential volume of offshore wind to be connected to the in the near future (Lee and Zhao, 2020). Ultimately, this has not turned out to be the case with record-breaking capacity additions in 2021 (World Forum Offshore Wind, 2021).

### 4.1.2 Analysis of outcome

Overall, it is evident that the winning bid price is relatively close to the price cap for the auction (i.e., CNY800/MWh). However, compared to the FIT price (CNY850/MWh) introduced by NDRC in 2014, there is a 6% to 11% decrease in the offshore wind energy price due to the introduction of the auction mechanism in 2019. Supplementary Note 2 provides the details on the observed bids the auctions conducted in China. The Chinese government is ambitious in reducing the cost of offshore wind energy and removing subsidies for new offshore wind projects in 2020 with subsidy-free projects expected in 2022 (Lee and Zhao, 2020).

It is conceivable that companies bid strategically to achieve cost reductions through economies of scale. Although unsuccessful bids are unknown, from the winning bids it is evident that several companies could win multiple projects auctioned at the same time. For instance, China Huaneng Group won three out of five projects auctioned in Wenzhou in November 2019, with the same bid price of 770 CNY/MWh. However, companies are still cautious about bidding extremely low prices, given the relatively low impact of price scores on the overall evaluation and the contribution of "enterprise capabilities" on the scoring. This includes investment capabilities, technical capabilities, qualification, and performance. It is also worth noting that the winning bids in several cases are relatively close to the price cap, which could indicate strategic bidding. Meanwhile, the design of the auction scoring ensures a relatively small difference in the scores of price criteria, which also reduces the price competition of bidders.

## 4.2 United Kingdom

### 4.2.1 Mechanisms

Offshore wind energy in the UK is supported using CfD auctions (BEIS, 2012), in which bidders compete on price for the CfDs with generation technologies at a comparable maturity level, and divided into distinct pots. The marginal pricing rule applies unless offshore wind sets the marginal bid. Each bidder can submit up to four hidden bids (i.e., not publicly disclosed) with different combinations for strike price, capacity, and delivery year (National Audit Office [NAO], 2018; Snape, 2016). The successful bidders and the government agree to a two-sided CfD for 15 years, which pays the generator any differences between the CfD strike price and the wholesale electricity market price from the CfD counterparty, the Low Carbon Contracts Company. CfDs are indexed to consumer price index increases since 2012. Costs resulting from the CfD to the Low Carbon Contracts Company are levied to electricity consumers. A new CfD round can be expected every year from 2023 onwards (Durakovic, 2022). The latest CfD auction for up to 12 GW opened in December 2021, with 6 GW specifically safeguarded for offshore wind ('Pot 3') (Durakovic, 2021, 2020a). The results for this are still pending in Q1 2022.

The failure to realise a project will results in the exclusion from any CfD auction in the following two years. Seabed leases must be acquired prior to CfD auction participation, which are tendered separately by the Crown Estate, with six auctions held to date. The current result from seabed leasing for 'Round 4' and 'ScotWind' will increase the total capacity to approximately 57 GW (The Crown Estate, 2021).

The cost of grid connection is borne by the generator/developer and priced into the CfD, whereas onshore integration is carried out by the grid operator. The developer can construct the connection and then sell the transmission asset to an (unbundled) offshore transmission owner, which charges for the grid use. Transmission costs were kept mostly constant for generators at £$_{2011}$10–12/MWh between 2006 and 2018, whereas distance to shore has increased ninefold (Offshore Wind Programme Board, 2016).

### 4.2.2 Analysis of outcome

Awarded bids have declined from £$_{2012}$114–120/MWh between the first round of CfDs in 2015 to £$_{2012}$37–42/MWh in the third round held in 2019 for delivery in 2024–25. More details on individual wind farms can be taken from our Supplementary Data 1 and Supplementary Note 3.

FID was made for 99.9% of the awarded capacity, the highest in any jurisdiction at the time of writing. The highly anticipated CfD results in 2022 will likely shift this number down, but do not change the fact that the UK's CfD scheme reliably delivers offshore wind capacity. Most of the projects take FID close to the award date; in one case on the date itself. Tight timelines of the CfD require the developments to pass the milestone requirement within 12 months (Maranca et al., 2017), either by spending 10% of capital expenditures or taking FID (Steven, 2017). Construction is underway or finalised for 2017 and most of 2019 projects (Buljan, 2021; Durakovic, 2020b, 2020c; Skopljak, 2020; SSE, 2020a). This strongly indicates that bidders have a well-understood cost basis while market price exposure is reduced by CfDs, creating certainty for investors and making delivery for the agreed CfD likely. We conclude that bid winners are anticipating the profitability of the wind farms.

Strategic bidding is a possibility for the 2017 CfD round. The auction rules are set to prevent strategic bidding but at the same time, market rules complicate bidding and reduce transparency (e.g., up to four bids, different delivery years, capacities, and prices). As a result, the delivery year 2021 has seen windfall profits for Triton Knoll, as the clearing price was likely to be set by fuelled generators and over the bid of the wind farm (Radov et al., 2017). Delivery year 2022 bids were lower, with only one bioenergy bid below the price of offshore wind not being able to increase the price. This is most likely due to the fuelled technology cap of 150 MW reached for both delivery years with 2021 bids (Radov et al., 2017). This has left the two wind farms in 2022 wanting for higher prices, which have not materialised.

It may be likely that bidders exercise some form of "option to build," in the absence of direct financial penalties (Welisch et al., 2019). Despite this, the measurable effect is small and can easily be overshadowed by 'overly optimistic expectations towards technology cost development', also known as the "Winner's Curse" (Welisch et al., 2019), so that it is unlikely that option bidding without the intent to build was executed on a larger scale. Furthermore, nondelivery would lead to an exclusion from auctions for 24 months, which means missing at least one CfD auction. This has happened on three occasions, but none of them being offshore wind farms (NAO, 2018). Delays in the delivery can cause CfD cancellation, posing a substantial financial risk, given the upfront investment needed for project development costs and acquiring the seabed lease costs from the Crown Estate. The transition from the more generous RO support to CfDs might have left many projects distressed, especially in less favourable conditions (Energy UK, 2018; Ofgem, 2018). However, the UK's ambitious target of 40 GW, technological advancement, and increased tender amounts make this less likely now.

## 4.3 Germany

### 4.3.1 Mechanisms

The German auctions for offshore wind energy are regulated by the Offshore Wind Energy Act (WindSeeG) and held by the Federal Network Agency (BNetzA). The WindSeeG, introduced in 2017, regulates the bidding procedure and coordinates the licensing, planning, construction, and commissioning of offshore wind and grid-connection projects. In WindSeeG's "transitional period," two auctions completed in April 2017 and April 2018 were held with predetermined commissioning dates between January 2021 and December 2025 for projects "at a very advanced stage of development" only. In total, 29 projects located in the North Sea and Baltic Sea met the requirements (Bundesamt für Seeschifffahrt und Hydrographie, 2018). In each auction, 1,550 MW of capacity were auctioned, with at least 500 MW reserved for the Baltic Sea and 60 MW of unassigned capacity from the first auction were rolled over to the second. Both auctions used the "pay-as-bid" mechanism.

The projects in both auctions competed for sliding market-premium payments, essentially a one-sided CfD, for 20 years from commissioning. This is shorter than the expected operational lifetime of 25 years, which could possibly be extended up to 30 years. Remuneration is based on nominal bids and not indexed to inflation. Bids exclude grid connection, which is provided by the grid operator and levied as a part of network charges (Deutscher Bundestag, 2020). All auctions are single-technology items for offshore wind only. However, alongside the current setup, Germany has announced combined offshore wind and hydrogen electrolysis auctions starting in 2022 (BMWi, 2021).

A modified version of the WindSeeG came into force by the end of 2020. It specifies the auction process from 2021 onward, when a so-called "central model" for offshore wind energy projects will be used for wind farms with a commissioning date after January 2026. In the "central model," preselection and investigation of the appropriate coastal sites will be performed by government authorities. The bidders will be invited to compete for the development rights of preselected sites in annual auctions. The auctions for the development rights on such sites will take place every September starting from 2021. In September 2021, the first three projects with a total capacity of 958 MW were auctioned, two in the North Sea (abbreviated N-3.7 and N-3.8) and one in the Baltic Sea (O-1.3). Penalties for non-compliance amount to €100/kW for existing projects (N-3.8 and O-1.3) and €200/kW for sites which have only been subject to a preliminary investigation (N-3.7) (Deutscher Bundestag, 2020).

In this context, "existing projects" are projects which were allowed to participate in the 2017/18 auctions – but did not get awarded. These projects are now auctioned under the new scheme – so every pre-qualified bidder may submit a bid for them. However, the previous developer got a right of subrogation for existing projects, i.e., the right to "step-in" at the terms of the winning bid at the site. Another feature of the auction format in Germany is that negative bids are not allowed. If several bids are submitted at zero, lots are drawn to determine a winner. There has not been an explicit seabed lease allocation in any of the German offshore wind auctions.

### 4.3.2 Analysis of outcome

In the first two German auctions, held in 2017 and in 2018, more than 50% of capacity awarded bid €0/MWh. In combination with a one-sided CfD, this implies that projects rely entirely on wholesale electricity market revenues (see detail Supplementary Note 4 for individual projects' auction results). This has attracted public attention in Germany and beyond, as commissioning of these projects implies economic competitiveness of offshore wind with conventional power generation. The implications of these bids are studied by Hübler et al. (2017) and Jansen et al. (2020).

The auction held in 2021 confirms this picture: All three auctioned areas were awarded at 0€ per MWh. Two areas (O-1.3 and N-3.8) even received multiple bids at zero. Hence, areas were awarded by lot. However, the "winners" of that lottery did not celebrate for long, as both areas were existing projects, and the previous developers could (and indeed did) exercise their right of subrogation and took over the projects (Iberdrola for O-1.3 with 300 MW and Northland Power with RWE AG for N-3.8 with 433 MW). The auction for N-3.7 was awarded to RWE AG with 225 MW. Such significant interest in these offshore wind projects from several developers even at 0€ per MWh is additional evidence of offshore wind's cost competitiveness (and the findings of Jansen et al. (2020). We thus discovered that investors expect project revenues on the wholesale market to be above project costs. At least three factors contribute to this: (1) project owners benefit from a long time span until projects must go online. For the first two auctions, realisation deadlines are established by the years of offshore grid connection, providing an opportunity for investors to make a FID later and benefit from longer learning intervals. Press releases (Güsewell et al., 2018; Innogy, 2018; Ørsted, 2018, 2017; Reuters, 2018) confirm that FIDs for winning projects were initially planned 2020–2023, with the majority of capacity now having taken FID (Ørsted, 2021). For the third auction, realisation deadline is 2026. (2) Project owners benefit from socialised grid-connection costs. Müsgens and Riepin (2018) evaluate the cost reduction for investors by looking at the shares of grid connection per kW installed in total project investment costs, which lie in a range of 15%–30%, with projects in the North Sea benefitting the most due to their distance to shore. (3) Project owners expect higher German wholesale electricity prices in the next decade. Ørsted (2017) names the phaseout of coal and nuclear generation capacities and a reinforcement of the European carbon trading scheme as key factors that will determine long-term prices, whereas the late realisation deadline allows for gathering market information before FIDs.

Caution is required when looking at the German results alone, as realisation of subsidy-free wind farms has yet to happen. First, a total of 29 projects with 8,654 MW aggregated capacity were eligible for the total auctioned capacity of 3,100 MW awarded in the "transitional period." Thus, only around one-third of capacity could receive permission to build a project with a market premium. In addition, the consequences of not winning were dire. Projects could even lose approval, so that building the project without subsidy became impossible. Hence, competition was fierce in these auctions, which may have triggered low bids and potentially contributed to the "Winner's Curse". Interestingly, competition between projects in the Baltic Sea was less severe because at least 500 MW was reserved for it—and higher bids in the Baltic Sea were awarded. Second, option bidding may have been present in these auctions. Müsgens and Riepin (2018) show that project owners pay a penalty of only approximately 3% of investment costs in the case of total nondelivery. Hence, the first German offshore auctions can be interpreted as auctioning an "option" to keep projects in a balance sheet and, when market uncertainties resolve, decide on further action. A late realisation deadline also makes such option bidding more attractive, although unlikely given that FID has now been taken from two thirds of the capacity.

The German auction design (both 2017/18 and current format) also has the special feature that winning projects receive the right for a grid connection to specific sea clusters (i.e., regions with physical connection points). The connection capacity of clusters is, however, limited. Thus, a bid is accepted "if it neither exceeds the auction volume nor triggers a capacity shortage within a cluster" (WindSeeG, 2017). Otherwise, the next (more expensive) bid is chosen. Müsgens and Riepin (2018) analyse the intracluster competition in the North Sea during the two German auctions and discover that cluster connection capacity is adding constraints beyond the overall auction setting, increasing competition even further. However, the authors report that there is no evidence for the more complicated strategy of blocking access in the first auction and reaping profits in the second.

## 4.4 Netherlands

### 4.4.1 Mechanisms

The Netherlands has organised offshore wind energy tenders since 2009, with a significant change in 2013 to the present "SDE+" competitive setup, as the result of an energy-climate agreement between government, industry, and stakeholders, culminating in the energy action plan. Offshore wind energy was given a substantial task in the agreement and the government set up a "Wind op Zee" (Wind at Sea) team. A reduction in costs of at least 40%, which was expected by 2020, was agreed upon compared to the prices of the Gemini and Luchterduinen wind farms. Six auctions were held up to 2020 in the SDE+ auction setup: three for the Borssele I-V area (1,400 MW), two for the Hollandse Kust Zuid I-IV (1,400 MW), and one tender for Hollandse Kust Noord V (700 MW), totalling 3.5 GW. The next, seventh auction, is in preparation and likely to mirror the setup of the last three auctions, which have resulted in subsidy-free projects. All auctions are for offshore wind only.

To reduce prospective developers' risks during the tendering phase, the government performs and provides site assessment campaigns, such as wind resource and met-ocean conditions and an initial environmental impact assessment, to all bidders free of charge. The Dutch transmission system operator (TSO) TenneT is responsible for developing and operating offshore wind energy substations and grid connections (Dutch Government, 2015) and guarantees that the nominal power of 350 MW (per site) can be exported to the main grid. Wind sites can be overplanted by 8% nominal power, though the operator is responsible to reduce power to the nominal power of 350 MW when instructed by TenneT. The cost for offshore transmission is socialised to accelerate development of sites. Site assessment costs, including a provisional environmental impact assessment, have been socialised to attract more relevant parties and lower bids. From 2022 onwards, (i.e. Hollandse Kust West (HKW) tender and later), the winner will have to reimburse the site development costs to the Dutch government, which is estimated at €13.5m for each HKW site.

Initially, the auction selection criterion was the cost of the energy. On bid submission, each bid bid must show a positive business case. On award, financial guarantees (i.e., bank guarantees) must be provided to ensure construction and commissioning within the agreed time. Without this, the government will reject the bid due to potential delivery risks, thus violating targets in the energy agreement. For the subsidy-free auctions, additional criteria (e.g., applied innovations) were implemented due to fact that cost of energy was not decisive anymore. Seabed leases are not awarded separately, but use of the seabed will incur charges. Non-delivery of a project could trigger a ministerial order for penalty payments to cover damages to the state and/or the revocation of the permit. A penalty of €10m must be paid if the bidder does not take the project forward, or fails to provide the bank guarantees, and incurs a penalty of €3.5m (up to ten times) for each month of delay in commissioning date (CMS, 2022).

Subsidies under the SDE+ system are a top-up (one-sided CfD, market premium) in addition to wholesale market price. A floor market price was determined of approximately €30/MWh, below which the risk falls on the developer and is not compensated by subsidy. The subsidy is paid for 15 years, starting from "first power" from the wind farm supplied to the grid, with no indexation to inflation. A maximum number of subsidised MWh per year is implemented by the auctioning body based on a state-of-the-art independent yield prediction report.

### 4.4.2 Analysis of outcome

The first two tenders under the SDE+ scheme for the Borssele I-IV wind farm was won by Dong Energy (now Ørsted) and Blauwwind, a consortium of Royal Dutch Shell, Van Oord, Eneco, and Mitsubishi/DGE. The prices of €72.7/MWh for the first and €54.5/MWh for the second tender were almost 50% lower than the maximum prices agreed upon in the energy action plan, which was a true surprise. Although subsidy-free offshore wind energy projects were not expected to be possible before 2030, it was decided to auction the first Hollandse Kust Zuid I-II tender without subsidies, which required changes to the tender award criteria from the lowest bid to other criteria. In case no eligible subsidy-free bids were received, a second tender round would be opened, based on the same criteria as for the Borssele I-V tenders.

The zero-subsidy bid included additional requirements to prove the validity of the business case with risks associated with fluctuating market prices. Some commentators referred to these tenders as a "beauty contest" (de Rijke et al., 2017), as some of the auction criteria are not standardised. The first zero-subsidy tender for the Hollandse Kust Zuid was won by Chinook CV, a subsidiary company of Vattenfall. The second tender for the Hollandse Kust Zuid III-IV organised in the first quarter of 2019 was also won by Vattenfall. The last tender in the road map to 2024 for Hollandse Kust Noord was organised in the first quarter of 2020, with a change in setup once again. Bidders were required to include a research program, paid by the project, on increasing the flexibility of the power characteristics of the wind farm in the electricity system. The Hollandse Kust Noord tender was won by Crosswind, a joint venture of Shell and Eneco, with a zero-subsidy bid. Inside the borders of the wind farm, part of the demo is to combine the wind farm with a floating solar farm (WindEurope, 2020), and a 200 MW electrolyser in Rotterdam (Radowitz, 2022).

For the parts of the Hollandse Kust Wind Farms that are located within the 12 miles zone, the owner pays a lease for seabed usage of ~€1m per year and wind farm; for Hollandse Kust III-IV, this is €2m annually in ground rent.

Each tender of 700 MW (e.g., Borssele I-II) is split into two adjacent sites of 350 MW each. Bidders that seek to combine both sites would need to bid and win with the lowest bid price for each site (Marsden and Radov, 2018). However, a combined bid allows for the argument that cost reductions due to economy of scale effects create a different business case. During the first tender for Borssele I-II, strategic bidding did occur, despite the government's efforts to prevent it through the tender design. The company winning the tender, Dong Energy (now Ørsted), entered 21 separate bids to ensure winning both sites at the same time. This loophole was closed in subsequent tenders, forbidding that equity of one parent company was used by more than two subsidiary companies. The Hollandse Kust Noord tender of 700 MW was organised as a single site.

The Dutch government is continuing its focus, with planned capacity additions of 1 GW per year in the period of 2024–2030 and a total capacity of 11.5 GW (Davis, 2016), with further fine tuning of the tenders to be expected.

## 4.5 Denmark

### 4.5.1 Mechanisms

In 2005, Denmark conducted the first competitive offshore wind energy auction for Horns Rev 2. Overall, nine auctions have yielded 3.2 GW of projects. The latest auction for the Thor project had final bid submission in late 2021. All auctions are individually agreed upon in a political process involving national parliament. The auction design is adapted for each auction and has seen considerable changes. Most auctions were single-item auctions (i.e., one offshore wind site) with predefined sizes of 200–1,000 MW. The nearshore areas were auctioned in 2016 as a multisite tender, with a total maximum capacity of 350 MW (Garzón

González and Kitzing, 2019). Access to seabed is granted without fees as part of the auction and permitting process.

The Danish Energy Agency (under the Ministry of Energy and Climate) is the auctioning body and grants licenses for preliminary studies, construction, and operation. Auction award criterion is the support level paid out as a two-sided CfD in øre/kWh of produced electricity, which is not indexed. From 2010 onward, and following the introduction of negative prices on the electricity market, no support is paid out in hours of negative spot prices. The nearshore auction had a price cap of €94/MWh, which is well above the realised result of €64/MWh (Danish Energy Agency [DEA], 2016). The 2021 auction had a cap on overall support being paid out over the lifetime of the project, as well as an overall lifetime cap on payments from operators to the state that occur in periods when market prices are above the CfD strike price (DEA, 2020).

The support duration is calculated for 55,000 full load hours over the project lifetime and is implemented in legislation as a specific amount of supported terawatt-hour for each wind farm individually (see Supplementary Data 1), which translates into expected support durations of 10–15 years (Kitzing et al., 2015). From the 2021 auction, support is paid for 20 years, not by full load hours (DEA, 2020), which makes Denmark a notable exception wherein the support duration has increased from around 12 years to 20 years, which aligns Denmark much more with other jurisdictions.

All auctions use static sealed-bids with pay-as-bid pricing. Typically, bidders must prequalify, apart from two early auctions. The Danish auctioning body conducts extensive stakeholder engagement. Several auctions contained a two-step process with a "first indicative offer" and a "best and final offer," between which tender design specifications were improved through individual meetings with the bidders (Held et al., 2014).

All offshore wind energy farms have guaranteed grid access, which is delivered to the offshore substation by the TSO Energinet. The nearshore farms are responsible for grid connection to shore. From 2021 onward, grid connection to the onshore substation is within the scope of the auction and constructed by project developers (DEA, 2020). Non-delivery of a project is subject to a lump-sum penalty, depending on the length of the delay, as well as in some cases a reduction of the amount of support (DEA, 2018).

### 4.5.2 Analysis of outcome

Only limited information is published by the auctioning body about bids and bidders. Prequalified bidder lists show that auctions attracted three to six bidders on average. The retake of the Rødsand II auction only saw two bidders and the Anholt auction only one. Information on unsuccessful final bids is not published.

The awarded strike prices were relatively low from the outset, starting from 51.8 øre/kWh (€69.5 MWh) for Horns Rev 2 (2005), then increasing slightly in 2008 and topping at 105.1 øre/kWh (€140.9 MWh) for Anholt (2009), then decreasing sharply until arriving at 37.2 øre/kWh (€49.9 MWh) for Kriegers Flak (2016). Here, it is notable that the nearshore bids were presented only 3 months prior to Kriegers Flak, resulting in 47.5 øre/kWh (€63.7/MWh) for Vesterhav Nord and Syd (2016), although comparatively lower cost could be expected due to shallower waters and shorter distance to shore. The high price in the Anholt auction has been justified by involved parties by the increased cost of wind turbines and installation equipment (especially of suitable vessels) in the period. It was investigated by a third party (Ernst & Young, 2010), which found the high price was partly due to supply chain bottlenecks caused by the simultaneous exponential growth of the early offshore wind energy sector across Europe. This and location-related issues, such as moving further offshore into deeper waters, can only partially account for the higher price of Anholt (2010), as compared to, e.g., Rødsand 2 wind farm (2008). Tight schedules, stringent penalty rules, and lack of competition might have contributed as well (Kitzing et al., 2015).

Profitability for developers seems to have been ensured by the auctions, as auctioned Danish offshore wind energy projects have been realised at the contracted sizes. The finding supports the notion that bidding has taken place with a firm intention to realisation, rather than "option bidding." Only the first auction of Rødsand 2 (2006) did not lead to realisation. The winning consortium (E.ON, Energi E2, and Ørsted) suggested a renegotiation of the price, and finally withdrew from the contract after failed efforts, so that the auction had to be repeated. They justified their withdrawal with heavily increased prices for wind turbines by the only two suppliers of large offshore turbines at the time (Vestas and Siemens) (Meister, 2007). This development can be an indication of the still limited maturity of the offshore wind energy market at that time. Also, underbidding may have taken place, considering that the winning party of the re-tender was one of the companies from the same consortium.

In the Anholt tender, only one bidder participated, so that the auction itself could not establish if competitive price determination had taken place. An analysis of the auction identified the main reasons for investors not taking part in the auction as: (1) the high penalties connected to delays combined with a tight schedule, and (2) the possibility of participating in offshore wind energy auctions on financially more attractive markets at the same time, especially in the UK (Deloitte, 2011). Furthermore, a clear policy for future Danish wind farms was lacking and so synergy effects to possible later projects were difficult to estimate. This is in stark contrast to the future Danish ambitions for offshore wind energy, where additional auctions of more than 15 GW are envisaged, which by far exceeds the country's current electricity demand and will contribute to the overall European transition to zero emissions (Danish Parliament, 2020).

The concession to build the Thor was won by German utility RWE AG after drawing of lots. This was necessary as five out of six bidders had offered the minimum price of 0.01 øre/kWh ("RWE wins Danish offshore wind tender Thor," n.d.). This bid price means that there will be no support paid out to the Thor project. On the contrary, the winner will have to submit the difference between the market reference price and the strike price of 0.01 øre/kWh (i.e. the full revenues), as payments to the Danish state during the first 2-3 years of operation, until the cap of DKK2.8bn is reached.

## 4.6 Taiwan

### 4.6.1 Mechanisms

Taiwan is the second-largest offshore wind energy market in the Asia-Pacific region. In January 2018, Taiwan increased its offshore wind energy target to 5.5 GW by 2025, with an additional 10 GW planned from 2026 to 2035 (Lee and Zhao, 2020). The capacity additions by 2025 were allocated through two different procedures: 3.8 GW through the so-called "selection procedure" and 1.7 GW through the "competitive bidding procedure" (i.e., auction rounds), which is also expected to be used for capacity additions up to 2035. The "selection procedure" allocated 3,836 MW to 11 offshore wind farms proposed by seven developers on 30 April 2018. Projects were awarded a 20-year FIT at NT$5850/MWh (€170/MWh). The selection criteria included technical (60%) and financial capabilities (40%). Local content consideration was an important factor in selecting the winning projects.

An auction process was introduced in mid-2018 to drive down the price. Taiwan's first offshore wind energy auction was launched on 22 June 2018, and allocated 1,664 MW to local and foreign developers through competitive bidding. Developers submitted sealed bids and the bidder with the lowest offtake tariff won. Unlike the "selection procedure," there was no local content requirement, and the main considering factor was the tariff. The final agreed-upon FIT is the lowest price between the granted price during the auction process and the announced FIT rate in the year of signing the power purchase agreements. New auction rounds are scheduled for 2021, 2022, and 2023 (Durakovic, 2020d), to fulfil Taiwan's 2035 offshore wind energy target. Their specific design is not available yet but is expected to be largely similar to the mid-2018 setup. Access to seabed is granted as part of the auction and consenting process. Offshore wind does not compete with any other technology for support.

### 4.6.2 Analysis of outcome

Five developers bid into Taiwan's first auction: 1) Northland Power & YuShan Energy, 2) Swancor & Macquarie, 3) Ørsted, 4) Copenhagen Infrastructure Fund, and 5) Taipower. Two sponsors, Northland Power & YuShan Energy and Ørsted, and four wind farms were selected, with a bidding price ranging from NT$2224.5/MWh (€65/MWh) to NT$2548.1/MWh (€75/MWh) and the average winning bid at NT$2386/MWh (around €63/MWh), marking an almost 60% reduction from the FIT awarded in 2018.

Taiwan's power grid system is run by the state-owned Taipower, which provides grid connection to all projects from the auction scheme. However, a commitment to timely connection is missing as well as compensation for possible delays, which is not covered by the FIT (Harris et al., 2019).

## 4.7 United States

### 4.7.1 Mechanisms

In the United States, development of offshore wind energy is governed by federal- and state-level policies. The issuing of seabed lease areas in federal waters is under the authority of the Bureau of Ocean Energy Management. To date, 17 lease sales were awarded through competitive auctions (Bureau of Ocean Energy Management, 2020). These provide the lessee with the rights to apply for and receive authorizations to assess, test, and produce renewable energy on a commercial scale over the long term. BOEM has wide-ranging authority over the auction mechanism and bidding process. For instance, auctions may be held in different bidding formats (sealed or ascending) and the award may be determined through single- or multi-factor bidding (Federal Register, 2022). The price for the lease awards has ranged widely between US$222 /km$^2$ (in 2015), US$261,872 /km$^2$ (in 2018), and most recently US$2,643,039 /km$^2$ per lease area (in 2022) (BOEM, 2022; Musial et al., 2019). Procurement of offshore wind energy is driven by individual states that have issued offshore-wind-specific procurement goals. This decentralised approach differs from European procurement, wherein the national government typically sets a uniform mechanism. Three offshore wind procurement pathways have been used by U.S. states to date: (1) PPAs with local utilities, (2) mandated ORECs contracts with state governments, and (3) utility-owned and operated generation (Beiter et al., 2020). Both PPA and OREC contracts provide a fixed price for the delivery of energy services. The fixed price guaranteed under a PPA or OREC tend to expose the contract parties to less risk of wholesale electricity price fluctuation—thus providing revenue stabilisation—than under a CfD remuneration that is used in many European countries.

In Massachusetts, Rhode Island, and Connecticut, state agencies have mandated local utilities to periodically issue solicitations for a predetermined capacity of offshore wind energy and to negotiate long-term (e.g., 20 year) PPAs for energy and environmental attributes. If contractually and technically eligible, offshore wind energy generators may access revenue streams outside of their long-term PPAs. For example, offshore wind farms may be able to participate in forward capacity markets. To date, five projects have executed 14 PPAs for 4,935 MW of total capacity (Musial et al., forthcoming).

In New Jersey, New York, and Maryland, state agencies competitively procure offshore wind capacity through long-term OREC contracts. An OREC represents the environmental attributes of one megawatt-hour of electric generation from an offshore wind energy project (Beiter et al., 2020). The OREC contract can either be fixed or indexed. Under a fixed-price OREC, the generator has a variable income from the wholesale market and fixed income from certificates. Under an indexed OREC, the generator receives a fixed income (see Supplementary Note 1 for details). Existing OREC contracts also enable offshore wind energy generators to receive revenue from participating in forward capacity markets. Currently, 11 projects have signed OREC agreements for 10,010 MW of total capacity (Musial et al., forthcoming).

The only utility-owned and operated project to date is the Coastal Virginia Offshore Wind project. This project is a demonstration project, with plans announced to extend to a 2.4-GW commercial-scale project by 2024.

### 4.7.2 Analysis of outcome

The PPA and OREC contracts negotiated between utilities and commercial-scale offshore wind energy projects have yielded prices between US$58/MWh (Mayflower Wind) to US$132/MWh (US Wind and Skipjack) (in levelised nominal terms and only for those with published price data) for a total capacity of 6.4 GW (Musial et al., 2019). Across U.S. states, expenses for the export system cable (and associated infrastructure, such as offshore substations) are the financial obligation of the offshore wind energy developer (and not the TSO, as in some European markets). States in the Northeast currently evaluate the merits and costs of a "backbone" system (Massachusetts Department of Energy, 2020). Offshore wind energy projects in the United States are expected to elect the federal investment tax credit. The credit was on a phase-down schedule and set to expire for projects starting construction after 2021 (Musial et al., 2020) but recent legislative action has introduced a standalone offshore wind investment tax credit at a rate of 30% that can be elected for projects with a construction start before 2026 (U.S. Congress, 2020).

When considering these unique features of competitive U.S. offshore wind energy procurement (Beiter et al., 2021), this first set of offshore wind procurements in the country seems within the range of recent Europe procurement prices. The prices might have been enabled, in part, by the fixed-price offtake regimes, industry confidence about the procurement of critical components, and historically low financing rates. Recent delays in permitting, however, have led several projects to push back their anticipated commercial operation date (e.g., Vineyard Wind).

## 4.8 France

### 4.8.1 Mechanisms

France National Regulatory Agency CRE (Commission de Régulation de l'Energie) started tendering for renewable projects in 2004 for a targeted capacity, using sealed bids and pay-as-bid pricing. Three offshore wind energy auctions occurred in 2011, 2013, and 2017. An earlier tender in 2004 was awarded but the projects were not pursued. The multiannual energy plan by the ministry in charge of energy defines the future tenders: (1) 1,000 MW in the East Channel (opened 2021); (2) 500–1,000 MW in 2021 in the South Atlantic and 250 MW in Southern Britanny for floating offshore wind; (3) 2 x 250 MW in 2022 in the Mediterranean for floating offshore wind; and (4) 1,000 MW in 2023 and 2024 each. The auctions are technology-specific (i.e., offshore wind only).

The first two tenders in 2011 (3 GW) and 2013 (1 GW) were awarded FITs with 20 years of support duration. All electricity produced is paid via power purchase obligations through EDF Obligation d'Achat, a regulated subsidiary of EDF, which is compensated directly from the state budget (République Française, 2016a). Auction bid prices should cover all costs related to the wind farm and the connection infrastructure. The price is indexed according to industry labour costs and production price indices and adjusted for annual full-load hours achieved and final capacity installed. The selection process is based on a multicriteria assessment weighting that includes (1) industrial evaluation by 40% (includes components production capacity, technical capacity, risk analysis and controls, or financial robustness); (2) care for existing activities and the environment by 20%; and (3) a bid price by 40%. The first and second tenders had floor prices in the range of €115-140/MWh and ceiling prices of €175-220/MWh.

The third tender call in 2017 (1 lot of 600 MW) modified the auction process by introducing a competitive dialogue phase and consultation before the final terms of reference were defined. The support scheme also changed to a two-sided CfD (République Française, 2016b), which is calculated as the product of the monthly electricity delivered times the difference between the strike price awarded and the average day-ahead market price in France (excluding negative price hours) weighted hourly by all wind production in France. In addition, all electricity retailers must purchase "capacity guarantees" from generators (e.g., wind, nuclear) to ensure that winter peak demand can be covered. The required amount of capacity guarantees is determined by the TSO, who also defines the amount of capacity guarantees that generators can sell. In the case of offshore wind energy, it is estimated that it will be able to sell 25% of its capacity as capacity guarantees, hence increasing its revenues. Finally, an annual capacity payment is deducted, which is calculated as the product of the number of capacity guarantees given to the wind farm times the arithmetic average of capacity market prices in the year before delivery.

Under this new support scheme, the offshore wind farm must sell its energy on the market and then receive monthly payments from EDF Obligation d'Achat. The support duration is 20 years. However, the connection approach applied in the third tender is a "shallow connection," thus all connection works are performed and covered by the TSO. The selection process adhered to the following criteria and weights: (1) bid price (70%), with a ceiling price of €90/MWh and no floor price, (2) contractual and financial conditions by 10%, (3) area occupied and distance to shore by 11%, and (4) the number of wind turbines and budget dedicated to environmental control by 9%. The weighting of the price weight has increased substantially from the first two auctions to the third, which is likely driven by the desire to reduce support costs (Lequien and Dabreteau, 2017). The third auction has also seen the addition of an offshore wind tax as an annual tariff of €17,227/MW installed, which is payable by the offshore wind farms. The bidder must acquire the seabed lease following an administrative investigation. The CfD duration can be reduced in case of later delivery by length of the delay for all projects following Dunkirk (Hogan Lovells, 2020).

The multiannual energy plan that defines the coming tenders in the next few years foresees decreasing ceiling prices in the auctions: from €60/MWh in the 2020–2021 tender to €50/MWh in the 2023 tender call (Transition and écologique et solidaire, 2020).

### 4.8.2 Analysis of outcome

Official documentation by CRE from the first tender in 2011, awarded in 2012, only presents the companies awarded for each of the lots auctioned and an estimation of the overall supports scheme costs, valued in €160/MWh (Commission de Régulation de L'Énergie, 2012, 2011). As for the second tender in 2013, awarded in 2014, no results have been published by the CRE (Commission de Régulation de L'Énergie, 2013). All six projects awarded in the first two auctions suffered delays in the permitting phase, due to lengthy administrative authorizations and appeals from third parties. In this context, in 2018, the government decided to renegotiate the tariffs initially awarded to account for substantial cost reductions observed in offshore wind energy projects and to avoid potential overcompensation (European Commission and Vestager, 2019). The new prices negotiated ranged from €131/MWh to €155/MWh, with a capacity-weighted average price of €140/MWh. Business plans calculated over these prices would allow project promoters to obtain an internal rate of return after taxes from 6.1% to 8% (European Commission and Vestager, 2019). Though originally part of the auction, these renegotiated FITs do not include the connection price component, which was moved to the TSO (a shift from "deep" to "shallow" connection). Additional modifications were made for price indexing and payments delays. As of 2021, three of the projects have initiated works for a scheduled commissioning from 2022 to 2024.

Results from the third "Dunkirk" offshore wind tender call in 2017 were awarded in 2019 (Commission de Régulation de L'Énergie, 2019). The winning bid was for a price of €44.0/MWh, and the range of bids from the eight prequalified bidders is €44.0/MWh to €61.0/MWh. As of 2020, the project is on the permitting stage and is due to be commissioned in 2027.

Although the difference in support mechanisms complicates the comparison and analysis of the evolution of prices along tender calls, a price reduction is perceived from the last agreed €140/MWh (average) in the first and second tenders and the €44/MWh awarded in the third tender call. Aside from general cost reductions in the offshore wind energy industry, the significantly lower price in the third tender can also be explained by more favourable conditions in the project site: softer seabed, higher wind speed, lower depth, and lower distance to shore.

## 4.9 Cross-jurisdiction results discussion

From the analysis of eight jurisdictions' auction schemes, we can conclude that policymakers make different decision for delivering offshore wind energy. The employed policy measures can be distinguished into two main categories: the way payment is allocated and additional policy support. We observe that the main focus of auction conditions is the price of electricity delivered, regardless of the form of payment (e.g., tax breaks, CfDs, FITs); China is an exception, though it still applies this criterion with less weight. Additional policy support outside the auction schemes (e.g., land lease agreements, grid connections, environmental studies, and others), are often granted to the wind farm developer at reduced or no cost. The auction setups also differ in the penalties applied for late or non-delivery, with some posing a significant financial risk, and less so from other schemes. Offshore wind capacity is auctioned in technology-specific setups only, without having to compete against other generation technologies. Regardless of the specific auction setup, each jurisdiction fosters competition between different players, as a common policy goal.

The auction schemes typically vary by the extent to which they expose project owners to power price fluctuation. Reduced price exposure is achieved, for instance, through feed-in tariffs with a guaranteed fixed price for electricity, whereas CfDs might only provide partial (or no) fixed price, depending on their design. Europe, with its longest track record on offshore wind energy, initially used fixed-price support regimes (e.g., renewables obligations and feed-in tariffs) with the stated goal to incentivize investment through de-risking (Kitzing et al., 2018) but has moved toward more risk-sharing support regime arrangements in some jurisdictions.

In Figure 6, we qualitatively assess the extent to which developers are exposed to market price risks. We use Beiter et al. (2020) for the classification, with the extension of considering support duration (with shorter duration increasing price exposure and vice versa). We correlate this assessment against market maturity, which is easily proxied by installed capacity, with markets over 1 GW installed considered to be mature (see Table 1).

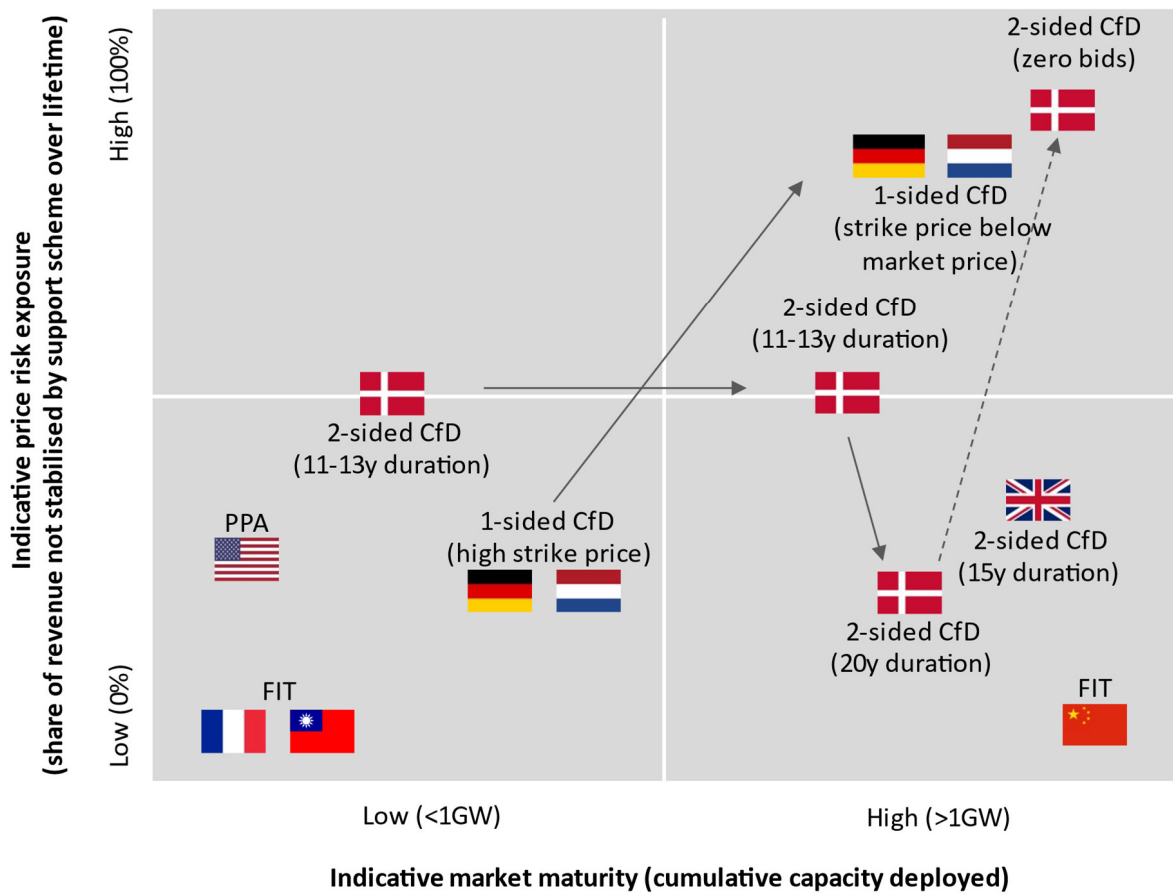

*Figure 6: Location of jurisdictions' auction schemes based on the markets' indicative maturity and price risk exposure managed by revenue stabilisation of these schemes. All markets with more than 1 GW installed are placed in the "high" area of the market maturity scale. Price risk exposure is qualitatively estimated based on Beiter et al. (2020).*

From Figure 6, we conclude that less mature markets tend to use revenue stabilisation measures more extensively. None of the immature markets chose to expose developers to market price risks on a large scale. For the mature markets, we observe a diverging trend between jurisdictions: Germany and Netherlands have moved toward higher exposures to market prices, whereas the UK, and China have opted to reduce the risk for developers. This is with the exception of zero-price bids, in which developers are fully exposed to market prices.

The case of Denmark is of particular interest in this context: On the one hand, Denmark has implemented a 2-sided CfD which socialises wholesale price risk. On the other hand, Denmark has capped payments resulting from the difference between bids and market prices in both directions (see 4.5). Depending on wholesale price distribution, price risk exposure for bidders can thus be either low (wholesale price is relatively close to bids) or high (payment cap is binding). In the recent Thor auction, bids of 0.01 øre/kWh in a 2-sided CfD thus indicate high risk wholesale price risk exposure for investors.

There is a clear trade-off for policymakers about the price exposure for developers. Early markets offer a lot to help incentivise and de-risk the technology, which is probably necessary to realise regional or national first-of-its-kind projects. Although developers in an established market may require less de-risking, high shares of wind have a greater impact on power prices, balancing, and grid stability. Policymakers may wish to expose wind farms to price signals through a risk-sharing approach to encourage better integration into the energy system.

From the increasing risk exposure and decreasing bid prices (down to zero-levels in some jurisdictions), we can deduct that offshore wind auctions face a transition from allocating support payments to awarding the right to build a (profitable) project. While this is an encouraging development and offers new perspectives for the decarbonisation of societies, it raises new questions for the design of auctions. For example, competitive bids could become negative as developers are increasingly willing to pay for the right to build. This can already be seen in recent seabed lease auctions in the US and the UK. Also, the Thor auction in Denmark showed that bidders are indeed willing to pay for the right to build on a merchant basis, even when support payments were available in form of a two-sided CfD.

Current auction designs prevent negative bids, which, in highly competitive situations can lead to multiple bids at the minimum allowed level (mostly zero), rendering a pure price-based selection mechanism ineffective. Already in two cases (in Germany and Denmark), auctioneers had to resort to drawing lots to identify an auction winner. In Germany, competing zero bids arose as the minimum allowed bids in the 2021 auction were at zero. Denmark's particular auction design for Thor has produced several competing zero bids, as the payment-capped two-sided CfD support mechanism turned the auction into a de-facto seabed lease mechanism, further blurring the lines between the two mechanisms. In the Netherlands, policy makers chose to expand the selection criteria and require additional capabilities from the bidders to avoid drawing lots.

These lotteries are far from ideal from a policy perspective, as this potentially means handing out large amounts of public funds to private entities that are selected by chance rather than capability, which may disincentivise developers from performing at their best. Furthermore, this runs the risk of realising less offshore wind capacity than developers would desire to realise at zero or very limited cost to society.

One solution to this policy dilemma may be to allow negative bids, which would see governments, taxpayers and/or electricity consumers realise additional profits. It does raise questions, though, of whether additional policies outside the auctions, such as seabed leases, scope of grid connection, and increasing requirements on environmental and social benefits, amongst others, should be reconsidered as well, to account for the new competitive market situation of offshore wind.

# 5 Conclusion and policy implications

We provide a comprehensive overview of auctions for offshore wind energy globally. Historically, offshore wind farms have predominantly been realised with administratively set prices, but future offshore wind energy is expected to almost exclusively be auctioned. We systematically describe the key features of auction designs and find notable diversity across the eight jurisdictions considered. The mechanisms in each jurisdiction have evolved and are embedded within their respective regulatory and market design context. Because of regional similarities in auction designs, we hypothesize that learning and spill-over might have occurred to some extent among neighbouring jurisdictions, which would be supported by earlier findings of Fitch-Roy (2015) on the converging European policy model.

All jurisdictions employ some form of a floating-for-fixed swap,[1] guaranteeing a fixed payment over a predetermined period of time or energy produced (Beiter et al., 2020). However, our results show that details matter. In designing support mechanisms, policymakers are confronted with a choice of risk allocation between private developers and a public entity (usually the rate payers) (Klessmann et al., 2008). In finding an adequate balance that sufficiently de-risks projects for developers while avoiding public exposure to disproportionate levels of risk, it has previously been argued that considerations of industry or technology maturity are highly relevant (see e.g., Kitzing and Mitchell, 2014).

Our results confirm this and add further insights: payment schemes tend to expose investors to less risk in less mature markets (e.g., the United States, Taiwan, France), whereas the situation is diverse in more mature markets. Some jurisdictions (e.g., the Netherlands, Germany) have opted for a higher risk exposure, essentially giving wind farms market revenues only, whereas others (e.g., UK, and China) have opted to limit the risk exposure for developers. For the latter group, the reduced risk for developers could lead to cheaper access to finance, hence overall lower project costs for investors. However, yet again details matter: Denmark has implemented a 2-sided CfD, thus reducing risk for developers. At the same time, Denmark has limited the CfD-payments which lead to investors taking significant price risk in the recent Thor auction. In any case, revenue stabilisation is employed for all offshore wind energy farms in addition to a broader policy environment, such as reliable electricity markets. In essence, policymakers actively choose the level of revenue stabilisation provided to best facilitate offshore wind energy deployment.

---

[1] A swap contract is a financial derivative in which two parties exchange cash flows from different financial instruments (Bartlett, 2019).

We identified bid limitations of zero (both in recent 1-sided CfD auctions in Germany and 2-sided CfD in Denmark) as potential design flaws as resulting lotteries are far from ideal from a policy perspective. We discuss allowing negative bids as a potential solution.

While bids cannot be compared to costs and across different jurisdictions, at least not without substantial harmonisation efforts, our data re-confirm the downward trend in costs across all jurisdictions described in Jansen et al. (2020) and Beiter et al. (2021). It also affirms the ability of bids from different auction designs to translate into cost reductions.

We observe that auction designs interact with broader policy design choices. The market situation is considerably different for developers when they must pay for their grid connection, and in particular when fees are to be paid for seabed access. Some countries conduct separate auctions for the allocation of seabed access, which undoubtedly has implications for the bidding behaviour in subsequent support auctions, in which bidding is especially distorted when competing projects pay different amounts for seabed leases. In auctions dedicated to offshore wind only, the separate execution of seabed and support auctions may lead to high prices paid for seabed leases, with an expectation that costs may be rolled over to the support auction. There is also an upcoming concern of increasingly high seabed lease costs, making offshore wind artificially less competitive compared to other low-carbon energy generation.

We conclude that auction designs may have implications for different levels of governance, as the future will see an increased momentum in offshore wind energy deployment. Auctions are clearly the preferred global choice of policymakers for realising more than 200 GW of offshore wind energy projects and to advance the decarbonisation of societies. The comprehensive overview provided in this analysis and insights derived from the detailed contextualisation of auction designs provide governments with the necessary information needed to deliver offshore wind through auctions, and ultimately, to enable the successful transition to a sustainable energy system.

# Author contributions

**Malte Jansen:** Conceptualisation; project administration; methodology; writing – original draft; writing – review and editing **Philipp Beiter:** Conceptualisation; writing – original draft writing – review and editing; **Iegor Riepin:** Conceptualisation; methodology; writing – original draft; writing – review and editing **Felix Müsgens:** Funding acquisition; conceptualisation; methodology; writing – original draft; writing – review and editing **Victor Juarez Guajardo-Fajardo:** Data curation; writing – original draft; visualization **Iain Staffell:** Funding acquisition; data curation; writing – original draft; visualization; writing – review and editing; **Bernard Bulder:** Validation; writing – original draft; data curation **Lena Kitzing:** Conceptualization; data curation; methodology; writing – original draft; writing – review and editing.

# Acknowledgements

MJ and IS acknowledge support from EPSRC via EP/R045518/1. IR acknowledges support from BMBF FKZ 03SFK5O0. FM acknowledges support from BMWK FKZ 03ET4067A. This work was authored [in part] by the National Renewable Energy Laboratory, operated by Alliance for Sustainable Energy, LLC, for the U.S. Department of Energy (DOE) under Contract No. DE-AC36-08GO28308. Funding provided by the U.S. Department of Energy Office of Energy Efficiency and Renewable Energy Wind Energy Technologies Office. The views expressed in the article do not necessarily represent the views of the DOE or the U.S. Government. The U.S. Government retains and the publisher, by accepting the article for publication, acknowledges that the U.S. Government retains a nonexclusive, paid-up, irrevocable, worldwide license to publish or reproduce the published form of this work, or allow others to do so, for U.S. Government purposes. The views expressed are purely those of the authors only.

This manuscript has been prepared based on the evidence only and is not intended as a comment on political issues.

# Data availability statement

All data collected for this paper are available in Supplementary Data 1 and can be found at DOI: 10.5281/zenodo.4672682 (https://doi.org/10.5281/zenodo.4672682), hosted at the Zenodo repository.

# 6 Supplementary Information

## 6.1 Supplementary Note 1: Definition of key auction design terms

- **Type of support mechanisms**

  The support mechanisms for offshore wind energy implemented worldwide show a great variation in their specific design. Payments differ between jurisdictions, and the specific design of the support mechanism gives rise to significant differences in the bids received, and thus must be accounted for when comparing auctions. We provide a systematic classification of support mechanisms in Table 3.

- **Support duration**

  The time period for which the support is granted. Support duration can be time-based (e.g., duration in years) or energy-based (e.g., fixed amount of produced electricity).

- **Market reference price**

  The basis for calculating the difference with the guaranteed price. This can vary in terms of spatial resolution (e.g., electricity price at the node where a project is connected or composite price index of a wider area) and temporal resolution (e.g., hourly price or averaged over a specific time period).

- **Inflation indexation**

  The adjustment of a guaranteed price to inflation; includes the choice of inflation index and base year for indexation.

- **Grid-connection costs allocation**

  The costs of connecting a project to the power system (e.g., transformers and substations, as well as the connection to the local distribution or transmission network). These costs can either be socialised (and thus considered as part of the grid infrastructure) or paid by a project developer (and thus considered as part of the wind farm).

- **Site development costs allocation**

  The cost of site selection and assessment can be allocated to a public body (e.g., federal or state government) or to a private entity (i.e., a project developer).

- **Penalty for noncompliance**

  The monetary and/or regulatory penalty imposed on the developer in case of noncompliance with project completion requirements (e.g., the project construction is postponed or cancelled). May include a partial or complete retainment of a security deposit (i.e., a bid bond in the form of a bank guarantee or cash deposit provided by each auction participant in advance).

- **Scarcity of auctions**

    The scarcity of auctions is discussed in light of (i) the ratio of auctioned capacity to the sum of capacities of all projects that participate in an auction, (ii) the consequences of not winning an auction, and (iii) the jurisdiction-specific aspects of scarcity (e.g., the German projects competed for limited grid-connection capacity in specific sea clusters).

*Table 3: Support mechanisms used in offshore wind energy markets, as summarised in Beiter et al. (Beiter et al., 2020)*

| | |
|---|---|
| Feed-In Tariff (FIT) | Developer receives a fixed price for each megawatt-hour (MWh) of electricity. Generator receives a fixed price for each megawatt-hour of electricity. Total remuneration is not exposed to changes in the commodity price. The fixed-rate tariff is typically set administratively with the auctioned quantity varying. |
| Power Purchase Agreement (PPA) or Renewable Energy Certificate (REC) (Fixed-Rate Instruments) | Developer receives a fixed-rate price from a contractual instrument (e.g., PPA or REC) or a regulatory order for a fixed quantity mandated by a government entity. Total remuneration is not exposed to changes in the commodity price. The fixed rate is set through competitive bidding (e.g., an auction). The commodity produced is usually sold to an intermediary (e.g., electric distribution company) who sells it into the wholesale market. |
| Contract for Difference (CfD), Two-Sided (Cap) | The generator receives the difference between the strike price and the reference price if the reference price is lower than the strike price. If the reference price exceeds the strike price, the generator does not retain the "upside" from the higher reference price but is required to pay it back to the administrator. The strike price is typically determined through competitive bidding (e.g., an auction) for a fixed quantity (or a budget) mandated by a government entity. Total remuneration is not exposed to changes in the commodity price. The electricity produced is typically sold directly into the wholesale market and receives the spot price. |
| Contract for Difference (CfD), One-Sided (Floor) | The generator receives the difference between the strike price and the reference price if the reference price is lower than the strike price. If the reference price exceeds the strike price, the generator retains the "upside" from the higher reference price. The strike price is typically determined through competitive bidding (e.g., an auction) for a fixed quantity (or a budget) mandated by a government entity. Total remuneration is exposed to changes in the commodity price only on the upside. The electricity produced is typically sold directly into the wholesale market. |
| Feed-In Premium (FIP) | The generator receives a fixed premium on top of the reference price. Total remuneration is exposed to changes in the commodity price. The fixed price premium is usually set administratively and the quantity varies. |
| PPA/REC (Fixed Premium) | The generator receives a fixed premium from a contractual instrument (e.g., PPA or OREC) or a regulatory order for a fixed quantity mandated by a government entity. Total remuneration is exposed to changes in the commodity price. The premium is typically determined through competitive bidding. The electricity produced is usually sold directly into the wholesale market and if applicable, environmental attributes (e.g., RECs) are sold to an intermediary (e.g., a distribution utility, state agency, or escrow account). |
| Quota | The generator receives a certificate price on top of the reference price. Typically, a government entity sets a target quantity and the price is determined in a certificate market. Total remuneration is exposed to changes in both commodity price and certificate price. |

## 6.2 Supplementary Figure 1: Awarded prices by jurisdictions

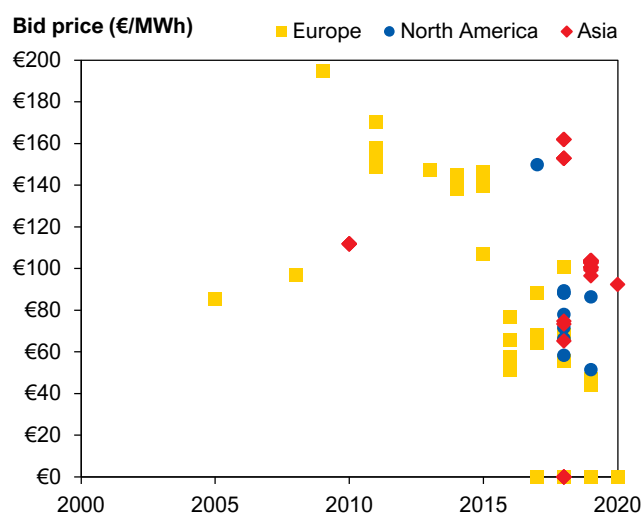

*Supplementary Figure 1: Awarded bid prices broken down by region.*

## 6.3 Supplementary Note 2: Observed bids in China

As mentioned in Section 4.1.1, while price weighs only 40% of the score in China's offshore wind auction, the lowest bid would not necessarily be the winning bid. The auction for the Fengxian project features a lowest bid of CNY620/MWh (€79.3/MWh), which was by China Longyuan, the wind development arm of the world's largest utility China Energy Investment Corp, which was well below the other bidders (between CNY738 and 760/MWh) and the winning bid (CNY738.8/MWh). For most offshore wind energy auctions in China that follow, not all the unsuccessful bids are published but the awarded bidders and their strike prices are usually published.

Two of the five projects in Wenzhou, namely Cangnan 1# and Cangnan 1# Phase 2, were won by China Resources with the winning bid of CNY785/MWh (€99/MWh), whereas the other three, Ruian 1#, Cangnan 4#, and Cangnan 4# Phase 2, were won by China Huaneng Group with the winning bid of 770 CNY/MWh (€97/MWh). The two projects in Ningbo, Xiangshan 1# Phase 2 and Xiangshan Tuci, were won by China Guodian Corporation and China Guangdong Nuclear Power Group with a winning bid of CNY760/MWh (€96/MWh) and CNY765/MWh (€97/MWh), respectively. Though won by different bidders, the four projects in Dalian, Liaoning province, have the same winning bid at CNY795/MWh. Similarly, the four projects in Yantai, Shandong province, also have the same winning bid price of CNY790/MWh (€100/MWh), though they are won by different developers.

## 6.4 Supplementary Note 3: Observed bids in the United Kingdom

The first round of CfDs in February 2015 awarded two offshore wind farms, 448 MW at Neart na Gaoithe in Scotland for £$_{2012}$114.39/MWh and 714 MW at East Anglia One for £$_{2012}$119.89/MWh (UKTI, 2015), now fully commissioned (ScottishPower Renewables, 2020). Neart na Gaoithe was held up due to a now-resolved environmental impacts dispute (Baosheng, 2018). The CfD had to be extended (BEIS, 2017), but the wind farm is now under construction (EDF, 2020) and expected to be fully commissioned in 2022/2023 (Baosheng, 2018).

The second-round CfD auction created headline-worthy low bids of £$_{2012}$57.50/MWh for the 1,386-MW Hornsea Project 2 and 950-MW Moray Offshore Windfarm (East), commissioned in 2022. The 880-MW Triton Knoll cleared at £$_{2012}$74.75/MWh for delivery in 2021. The bid for Triton Knoll was likely to be lower than the awarded CfD, as it appears that the clearing price was set by fuelled technologies for delivery in 2021 (KPMG, 2017). This was not case for projects with delivery in 2022, as the only awarded non-offshore bid was below the awarded CfD bids. Individual actors indicated that at least one bid was lower than £$_{2012}$57.50/MWh but ultimately was excluded due to exceeding the capacity limit of 1.5 GW imposed on the bids (NAO, 2018). All three projects will be built in three phases, wherein stated delivery dates are the commissioning dates of the first phase.

The third auction saw bids reduced by another 30% with 2.6 GW of capacity with delivery in 2024 at an awarded a strike price of £$_{2012}$39.65/MWh and 2.9 GW at a strike price £$_{2012}$41.61/MWh for delivery in 2025. A cluster of 3.6 GW at Dogger Bank that has been won by a joint venture of SSE and Equinor will be the world's largest wind farm. It is currently under construction after financial close (FID) in November 2020 and ordering of the GE Haliade-X 13-MW wind turbine in September for two-thirds of the project (Foxwell, 2020; Paulsson, 2020) and the 14-MW version of the Haliade-X for the other third (SSE, 2020b). The fourth auction round is currently underway, and results have yet to be published.

## 6.5 Supplementary Note 4: Observed bids in Germany

Table 4 summarises the results of the three German offshore wind energy auctions. The results of the first German auction in April 2017 stunned the energy industry, with three out of four winning projects commissioned in the North Sea bidding €0/MWh. The average volume-weighted auction price for 1,490 MW of awarded capacity was just €4.6/MWh. The second auction took place in April 2018. There were six winning projects—three in the North Sea and three in Baltic Sea. The average volume-weighted auction price for the awarded 1,610 MW was €46.6/MWh. The strike prices varied significantly between the highest bid of €98.3/MWh and €0/MWh. The third auction, under a revised auction scheme, took place in September 2021. Three different areas were auctioned. All three were awarded at 0€ per MWh. Two areas (O-1.3 and N-3.8) even received multiple bids at zero.

Information provided by the auctioning body (BNetzA) includes the names of winning companies and the highest/lowest and average bids, as well as the total capacity auctioned. Though not published by BNetzA, project-specific data on bids can be obtained from companies' public press releases.

*Table 4: Winning projects in German auctions for offshore wind energy*

| Project Name | Current Owner (2021) | Capacity Awarded (MW) | Raw Bid (EUR/MWh at Award Time) | Award Date | FID | Operation Start (Funding Start) | Construction Start |
|---|---|---|---|---|---|---|---|
| OWP West* | Ørsted | 240 | 0 | 04/2017 | 2021 | 2024 | - |
| Borkum Riffgrund West II* | Ørsted | 240 | 0 | 04/2017 | 2021 | 2024 | - |
| Gode Wind 3 | Ørsted | 110 | 60 | 04/2017 | 2021 | 2023 | - |
| He Dreiht | EnBW AG | 900 | 0 | 04/2017 | exp. 2023 | 2025 | - |
| Baltic Eagle | Iberdola | 476 | 64.6 | 04/2018 | 2020 | 2022 | - |
| Gode Wind 4** | Ørsted | 131.75 | 98.3 | 04/2018 | 2021 | 2023 | - |
| Wikinger Süd | Iberdola | 10 | 0 | 04/2018 | exp. 2020 | 2022 | - |
| Kaskasi | Innogy | 325 | - | 04/2018 | 2020 | 2022 | - |
| Arcadis Ost 1 | KNK Wind5 | 247.25 | - | 04/2018 | - | 2023 | - |
| Borkum Riffgrund West I* | Ørsted | 420 | 0 | 04/2018 | 2021 | 2024 | - |
| O-1.3 | Iberdrola | 300 | 0 | 09/2021 | - | 2026 | - |
| N-3.7 | RWE AG | 225 | 0 | 09/2021 | - | 2026 | - |
| N-3.8 | Northland Power and RWE AG | 433 | 0 | 09/2021 | - | 2026 | - |

\* Now part of Borkum Riffgrund 3

\*\* Now part of Gode Wind 3